\documentclass[12pt,a4paper]{memoir}
\usepackage[T1,T2A]{fontenc}
\usepackage[utf8]{inputenc}
\usepackage[greek.ancient,latin,german,french,dutch,italian,spanish,russian,english]{babel}
\usepackage{amsmath}
\usepackage{amsfonts}
\usepackage{amssymb}
\usepackage{makecell}
\usepackage{amssymb,amsthm,enumitem}

\DisemulatePackage{setspace}
\usepackage[onehalfspacing]{setspace}

\usepackage[backend=biber,style=verbose-trad1,language=auto,autolang=hyphen,autocite=footnote,citetracker=true,ibidtracker=false,idemtracker=false,opcittracker=false,citereset=chapter,citepages=omit,isbn=false,url=false,doi=false,eprint=false,ibidpage=true]{biblatex}
\DeclareUnicodeCharacter{2212}{-}

\usepackage[style=english]{csquotes}
\usepackage{calc}
\usepackage{textcomp}
\usepackage{hyphenat}
\usepackage{keyval}
\usepackage{ifthen}
\usepackage{url}
\usepackage{etoolbox}
\usepackage{nicefrac}
\usepackage{amsmath}

\emergencystretch=\maxdimen
\AtEveryBibitem{\clearfield{note}}
\AtEveryCitekey{\clearfield{note}}
\AtEveryBibitem{\clearfield{isbn}}
\AtEveryCitekey{\clearfield{isbn}}
\AtEveryBibitem{\clearfield{url}}
\AtEveryCitekey{\clearfield{url}}
\AtEveryBibitem{\clearfield{lccn}}
\AtEveryCitekey{\clearfield{lccn}}
\AtEveryBibitem{\clearfield{doi}}
\AtEveryCitekey{\clearfield{doi}}
\AtEveryBibitem{\clearfield{eprint}}
\AtEveryCitekey{\clearfield{eprint}}
\AtEveryBibitem{\clearfield{issn}}
\AtEveryCitekey{\clearfield{issn}}
\AtEveryBibitem{\clearlist{publisher}}
\AtEveryCitekey{\clearlist{publisher}}
\AtEveryBibitem{\clearlist{language}}
\AtEveryCitekey{\clearlist{language}}
\AtEveryBibitem{\clearfield{note}}
\AtEveryCitekey{\clearfield{note}}
\usepackage{hyphenat}
\setsecnumdepth{section}
\setsecheadstyle{\large\textbf\centering}
\setsubsecheadstyle{\large\centering}
\usepackage[hang]{footmisc}
\usepackage[twoside=false,bindingoffset=1cm,left=2cm,right=1cm,top=2.5cm,bottom=2cm]{geometry}
\usepackage{calc}
\usepackage{textcomp}
\usepackage{microtype}

\DefineBibliographyStrings{english}{
	pages = {p\adddot}}
\DefineBibliographyStrings{french}{%
	pages = {p\adddot}}

\usepackage{stringenc}
\usepackage{pdfescape}

\DeclareFieldFormat[book]{shorttitle}{{#1}}

\DeclareNameAlias{sortname}{last-first}
\DeclareNameAlias{default}{last-first}

\DeclareSourcemap{
  \maps{
    \map{
      \step[fieldsource=language, fieldset=langid, origfieldval, final]
      \step[fieldset=language, null]
    }
  }
}

\DeclareSourcemap{
	\maps[datatype=bibtex]{
		\map{
			\step[fieldset=hyphenation, fieldvalue={english}]
		}
	}
}

\author{Starostin D. N.}

\title{Astronomical Cycles and Late Antique Chronology}

\date{}

\setsecnumdepth{subsubsection}
\counterwithout{section}{chapter}

\begin{document}

\maketitle

\author{Starostin, D. N.}

\begin{abstract} This article advances the hypothesis that the heightened eschatological sensitivity evident among the historians writing in the 5th century and its weaker echos in the time of Charlemagne were caused by the irregularities of the the lunisolar calendar and its particular realization, the Easter calendar. The lunisolar calendar that Christians used for the calculation of the date of the Easter had a number of key periods when the cycles of the Sun and the Moon came in sync in relationship to the beginning of the count and thus produced an effect of the times repeating themselves or ending with the nearly precise astronomical repetition. In this article several key lunar cycles that stemmed from the harmonics of the Moon’s precession on its orbit around the Earth are outlined on the basis of the modern astronomical data and with the help of simple mathematical calculations. It is suggested that the conjunctions of the solar and lunar calendars fell on the 1st and on the late 4th and 5th centuries CE, with a long hiatus in the 2nd and  3rd centuries. It is argued, at the same time, that there were significant irregularities in the lunisolar calendar that were visible in the 5th century and ca. year 800~CE. These irregularities when the calendar either lost or added a day due to the imperfections of the Julian calendar and to the lack of knowledge about the true length of the solar and lunar periods may have contributed in the 5th century and ca. year 800~CE  to the heightened expectations of the time (or the lunisolar cycle) reaching its end in disarray. It is suggested that the first episode of facing off with the irregularities of the lunar calendar of Easter holidays came at about 410~CE when the problem did make historians wonder whether the Heavens all went against the normal course of time. Thus sensitivity to eschatological vision of time in the 5th century and ca. 800  CE may be hypothetically explained by the fact that the solar calendar stopped syncing with the lunar calendar and discrepancies arose.

\end{abstract}

\mbox{}

\section{Introduction}

An investigation of how historical writing and chronological schemes were constructed in Late Antiquity needs to take into account the astronomical and mathematical foundations that underlay the structures of narratives about the past. In this paper I seek to prove a hypothesis that the emergence of Christian histories in the wake of St.Augustin’s writing the “City of God” was due to the predictions that the astronomers and specialists in calendric matters had learned to make while the Christian vision of history and the practical methods of calculating the date of Easter were gaining momentum. I argue that to understand the interest to the eschatological matters that had appeared in the 5th century in the histories of Orosius, Sulpicius Severus, Prosper of Aquitaine and Hydatius one first needs to investigate the spectrum of astronomical knowledge that was available to computists and historians when they had to do the calculations of time. In the following section I will first shortly address the groundbreaking lunar theory of the day Enlightenment Swiss 18th-century scholar Leonard Euler in order to show that the Easter calendar of Late Antiquity lay on a solid astronomical foundation. I will then outline the width of astronomical knowledge about the cycles of the Sun and the Moon that was available to scholars of that age. Finally, I will introduce a new method of discussing the relevance of Late Antique astronomical knowledge to the Christians seeking to establish a stable routine to remember and celebrate the Resurrection. In particular, I will pay attention to whether the writing of history was influenced in any way by the long cycles of the lunar calendar and, especially, by the repeating patterns of the Moon showing on the same day and time of the solar year over periods much longer than the length of human life. This will be the special approach that I seek to test on Late Antiquity’s representations of history. I will outline a series of dates in the period between the birth of Christ and the emergence of the Carolingian Empire that were the time when the Moon's motions came to be synchronized with those astronomers could see at the year of the birth of Christ. I will draw attention to one particular lunisolar cycle of 483 years that was mentioned in the Book of Daniel 9:25 and that may have contributed to astronomers and historians becoming particularly aware of the “end of times” in the 5th century CE. In other words, in the following discussion I will argue that the eschatological sensitivity was heightened in the 5th century by a number of mathematically-determined trends in the Julian/Christian calendar that made scholars speak about ending one cycle of the history of the Mediterranean and restarting the calendar to counter the chaotic effects of time (or particularly, of the Moon) running out of sync with the Sun and the stars.

Calculations in this paper are undertaken as a very approximate representation of the lunar Motion around the Earth. With the basic formulae that rely on publically available data related to the astronomical constants, the aim of thi paper was to make a first iteration to the problem of how relevant astronomical calendar was for people in Late Antiquity. A large number of cases in which the synchronism of the solar and lunar calendars fell on the dates when important political events took place suggests that this very approximate method can bring interesting results. Scholars who are currently calculating the parameters of the Moon's motion in the past with the help of J. Meeus' sets of algorithms may correct my results in the future.\autocites{Meeus:1998} It is important, however, that a very basic approach provides the dates which every hisotry of the Anciet world and the Middle Ages easily recognizes as critical.

\section{Historiography}

Modern scholarship has shown how during Late Antiquity Christian history-writing was forged out of the traditional Roman historiography and how drastically it changed its paradigm. Eusebius merged the chronological systems of the different cultures of the Mediterranean into one coherent system and, most importantly, into one tabular representation. But even after this work the question of constructing a coherent narrative remained unfinished, since the former did not provide a link between the eschatological motifs of the Old Testament past with the conflicting current history of the Roman empire in a long-term eschatological chronology. This task, started by Jerome’s commentary of Daniel and his interpretation of the idea of four world empires in order to provide a Christian model of history that connects the Ancient world with the Roman Empire. The writing of Eusebius, Jerome, Sozomenos, Socrates and Philostorgios sought to unite in one timeline the history of the Ancient empires with the one presented in the Old Testament.  Theologians like Augustine and the historians who followed in his path (Orosius, Hydatius, Sulpicius Severus, Prosper of Aquitaine) contributed to writing a new historical narrative that imposed the Christian vision of history, with its resonant eschatological theme, on the histories of the Roman empire and the regions comprising it. It has been argued that the biblical ideas of how history should be written intertwined with traditional views of the past into a peculiar Late Antique amalgam.\autocites{Burgess:1999}{BurgessKulikowski:2013}{Burgess:2012}{Burgess:2012a}{Zecchini:2003}{Croke:1983}{Croke:1987}{Croke:2005} By the 5th century, scholars have shown, the historical narrative gradually shifted from the chronological frame of Roman history ab urbe condita to the Christian time frame that built upon two main events: the Creation and that of the spread of Christianity.

The work of Richard Landes advanced the study of Christian chronography and chronology and drew the attention of scholars to the fluidity of the calendar when it was used by Late Antique chronographers to calculate the “end of times”. During Late Antiquity it was taken as the basic belief that one period of human history would end 6,000 years from the Creation. During the 4th- and 5th-centuries the birth of Christ was believed to have taken place in either 5,618 (Orosius), in 5,492 or in 5,508 years from the Creation. During the 5th century (with its expectations of the end of the world) chronologists had the need to recalculate and readjust the calendars so they gave more time to the society —--both ecclesiastical and lay —--before the final countdown. R. Landes noticed that over the course of the 5th and 6th centuries, the date given by Orosius of the birth of Christ (5,618) was readjusted to the one given by Gregory of Tours (5,199). He argued that this shift showed how deep the eschatological thinking was running in the cultural matrix of that age. This adjustment allowed Late Antique historians either to avoid acknowledging the end of time during their lifetimes or explain why nothing extraordinary had happened when the end of times was to come. The decision to shift from the year 6,000 from the Creation to a later date helped Richard Landes to argue that eschatology was a crucial element in thinking about history among Christian philosophers.\autocites{Landes:1992}{Landes:1993}{Landes:2000}

Astronomical knowledge was limited to a small community of the educated and was largely irrelevant for the masses. These concepts that deal with the eschatological sensitivity must rely on the works that showed how the visions of the orderly community, emerging from the writings of St. Augustine and other church fathers, were picked up by the converts and believers from their pious leaders. The importance of the holy men who sought to instill the Christian values and the way life has been shown by scholars. \autocites{Brown:1982}{Brown:1992}{Brown:1981}{Brown:1971a}{Brown:1996}{Brown:1971} But one must not forget also that the value of the Christian calendar was delivered to the masses by bishops, who in addition to the ascetics, sought to show the correct way of adapting the cycles of life to the cycles of the lunisolar calendar.\autocites{Pietri:1983}[279-300]{VanDam:1985} Thus this study seeks to investigate how important was the sacred knowledge of astronomical cycles, which were especially relevant for the leaders of the Christian communities, for the masses. The repetitions in the lunisolar cycle were important for historians and the advisors of rulers, but I suggest that leaders of the Christian communities, bishops and presbyters, sought to relate by their example the importance of the critical events in the universal Christian history whose time was calculated with due heed of the astronomical cycles.

The work of the American historian allowed further discussion of medieval eschatology and revitalized a topic that used to be overlooked in the past since it had been considered to be, since the coming of the 19th century, the reflection of outdated beliefs of the clergy. The work of Richard Landes allowed further to understand Late Antique chronologies and systems of time reckoning built on the principle of linearity, the attention to which had been drawn by A.~Momigliano. Since the “end of times” did matter, as Richard Landes sought to prove, that meant that the understanding of history as events that ought to be placed in linear fashion —-- with a clear beginning and end —-- took hold by the time of Gregory of Tours.\autocites[100-103]{Heil:2000}[130]{Palmer:2014}[198]{Reimitz:2004}{Borst:1972}{Borst:1992}[51]{Warntjes:2018}[381]{Nelson:2019} His publications on the subject were also employed to emphasize the demand for a linear construction of Christian chronologies the 4th and 5th century. They also contributed to illuminating the existential crisis experienced by historians and theologians when eschatological thinking came to dominate their choice of dates and critical events in their narratives. In other words, if his assessment of the importance of eschatology among historians and theologians was true, then it was also true that the was indeed an close relationship between the development of historiography in this period and the existential crisis provoked by the need of an imminent end in the development of Christianity.

It is well known that Late Antiquity was the period of re-writing or re-arranging history with the purpose of reconciling the era of the Creation according to the Bible and the Christian vision with the calendar of the Roman Republic and the Roman Empire. In the 3rd century scholars like Tatianus of Antioch (fl. 180), Clement of Alexandria (before 215), Hippolitus of Rome (died 235), Julius Africanus (200-245), Eusebius of Caesarea and Pseudo-Justin put a lot of effort in setting the Old Testament chronologies, frequently citing the previous Graeco-Jewish biblical chronographers, to determine the correct number of years that extended between the Creation to the Incarnation of Christ.\autocites{Africanus:2007}[451-452]{Wacholder:1968}{Gelzer:1978}{Mosshammer:2006} But let us look first whether there was some reason for raising eschatological concerns in the 5th century and ca. 800 CE in the peculiarities of the lunisolar calendar.

\section{The Cycles of the Earth-Moon System and its Relevance for Late Antique Chronology}

\subsection{The Cycles Known to Scholars In Late Antiquity}

A key question is why the eschatological expectations ran high in the 5th century, since the astronomical calendar is known for its perennial stability. Was it just the cultural phenomenon, exacerbated by the social and political contexts? Was there anything special about the 5th century in terms of the lunisolar calendar?

The longer cycles of time were still in the process of being discovered as Hypparchus established by 127 BCE that the equinoctial and solstitial points were moving against the fixed stars by no less than 1 degree per century.\autocites{Kugler:1900} But there was another cycle built into the motion of the Moon. The repetition of the Moon’s phases is the synodic month of 29.53 days (an approximate value). But since the Moon’s orbital plane is inclined to the ecliptic plane at 5.14°, the Moon is seen going seen going through the plane of the ecliptic into the Northern Hemisphere or into the Southern Hemisphere every 27.212 days (an approximate value). At the end of 345 and 578 years the phases of the Moon according to the synodic months coincided with the position of the Moon against its node, that is, with the count of draconic months.\autocites[122]{Aaboe:1955} For those reading Ptolemy these cycles were the basic element of astronomical and calendric knowledge and that influenced the worldview of specialists in chronology.

Before the introduction of the Christian calendar the astronomical period of about 72 to 76 years had been discovered by Hypparchos (c. 190 –- c. 120 BCE), who noticed the precession of the Earth against the fixed stars by observing the sky at the time of Vernal equinox. This allowed Hypparchos to observe that the Moon took the same synodic position on time while it took Earth more time to rotate so that the stars would take their preciously observed synodic positions.\autocites[For the purposes of this paper we not consider the secular variation in the speed of the rotattion of the Earth that has been calculated by:][]{Morrison:Stephenson:2004} This precession amounted to about a degree in approximately 72 or 76 years. He is known to have introduced into Greek science a “large cubit” of 2 or 2.5 degrees, which he had borrowed from Babylonian scholars.\autocites[104]{Bowen:Goldstein:1991} This period is known as {``}Callyptic cycle{''}, named after Greek astronomer Callyppus (ca.~370~BCE -- ca.~300~BCE), and it is remarkable because after 76 year the Moon's phases return to the same day and hour of the solar calendar if one day is omitted .\autocites{Kieffer:Callyppus} Leonard Euler was able to find correct mathematical equations for this phenomenon and proved its compliance with the fundamental Newton’s law of gravitation and the 2nd law of Kepler.\autocites[317]{Euler:E187:1753} Moreover, this period of near total conjunction of the Moon phase on the Vernal equinox of the year 1 BCE with the phase of the Vernal equinox each 76 or so years is another expression of the phenomenon of the Earth’s precession of 76 years against the fixed stars that Leonard Euler found in his calculations in the 18th century.\autocites[296--336]{Euler:E223:1756:Variation:de:latitude}[583]{Verdun:2015} In addition, modern astronomical measurements show that the Moon latitude’s free libration is elliptical with the axes of 3'' by 8'' with the period of 74.6 years, which is quite close to 76 years.\autocites[43]{Newhall:Standish:1996} In other words, in about 76 years many astronomical phenomena can be observed at once. Euler’s achievement was in calculating with mathematical methods the exact value of this precession which coincided with the observed phenomenon.

\subsection{The Cycles Calculated by L.~Euler}

One of the key methods in the approach I propose to employ in the investigation of the solar and lunar cycles is to make renewed emphasis on the work of Leonard Euler (1707--1783) in the field of the theory of the Moon's motion. The reason for this emphasis is that the starting point of his work, as we have seen, was finding a mathematical proof for the Callyptic cycle. In his works on the motion of the Moon Leonard Euler seemed to pay attention to the astronomical knowledge of the past that had been accumulated in the Alphonsine and Rudolphine astronomical tables. This is why he was one of the scholars who, in the Age of Enlightenment, helped build the bridge between the astronomical perceptions of Antiquity and the Age of Modernity. Let us pay attention to the other lunar cycles, which emerge from his calculations, that seem to correlate with the Late Antique practices.

\begin{enumerate}

\item One may use for further discussion the best of astronomical and mathematical achievements of the Enlightenment Age, the works of Leonard Euler (1707–1783). The reason for turning to his contribution to astronomy lie in the fact that he was the person who seemed to navigate the wealth of knowledge coming from the Alphonsine and Rudolphine astronomical tables (perhaps, through second-hand scholars) because he was putting the question about the motion of the Moon in the same fashion it was have been put by a scholar trained in Ptolemy’s and medieval astronomy. L. Euler has determined the precession of the apogee of the Moon to be either 113.7 or 117.2 years, which correlates with the 114-years Easter cycle.\autocites[][119, §136; 177, §204. He calculated a coefficient of the Moon's displacement to be equal to 1.0087947 in contrast to the earlier calculated values of 1.0085272. The difference with the modern value of 74.6 years comes from the fact that this latter value is based on the different set of parameters that are modern calculations because they come from modern space laser distancing measurements and it uses the system of coordinates different from that which L.~Euler was using in his terrestrial visual observations.]{Euler:E187:1753} Let us note that a period of this length (112 years) consists of seven 16-year periods that were used in Greek calendars to establish synchronicity between the solar and the lunar calendars because in 16 years the Full Moon comes 3 days later in the days of the Solar calendar than in the initial year. In seven cycles of 16 years the Full Moon was easy to predict because it would fall on the same day of the week. It may also be noted that the difference between the sidereal year of 365.256 days and the year of 365 days, 5 hours, 55 minutes, and 58 seconds (365.2472) calculated by Erasmus Reingold ca. 1551 also produces one-day ddifference in 112 years. This suggests that L.~Euler may have been looked to establish his calculations on the Julian and heuristic medieval calendar that had long served his predecessors in the field of astronomy.

\item L.~Euler also found mathematical proof for the cycle that, as he believed, determined the precession of the distance between the Earth and the Moon to be about 343 years.\autocites[175–176, §203]{Euler:E187:1753} This does not correspond to modern theory, but this relates well to the cycle of 345 years, after which the anomalistic and the draconic months converged.\autocites{Rawlins:2002} This meant that after 345 years the Moon appeared to be of the same size and at the same inclination to the horizon.

\item He also found that the cycle for the ascending node of the Moon was 235 years.\autocites[182, §210]{Euler:E187:1753} This would not be correct according to the modern data, but this period was observed by Anciend astronomers and therefor was important as part of cultural practices of that age. Let us note that although people in the ancient world and in Late Antiquity could not calculate these periods with precision, these cycles had distant echoes in the empirical chronologies they used.

\end{enumerate}

\subsection{The Conjunctions of the Solar and Lunar Calendars}

\subsubsection{The Conjunctions of the Astronomical Solar and Lunar Calendars}

I further suggest that there are other astronomical cycles that could have been observed by Christians, scholars and theologians of Late Antiquity. In particular, the 5th century as counted from the birth of Christ did have a number peculiarities in the astronomical sense and in the sense of the mathematical underpinnings of the calendar. In light of the new advances in the astronomy I would attempt to draw attention to one particular coincidence that may theoretically serve as an explanation of the 5th-century’ particular disquietude in the writing of history. Because the astronomers of Late Antiquity may not have known that the actual length of the year was 365.24219 days and that the lunar month was actually 29.53 solar days (29½ days plus ca. 43 minutes), there accumulated in their calendars a small but significant discrepancy.\autocites{Newcomb:1898}{Dershowitz:2008}{Meeus:Savoie:1992}{Meeus:2009} Both the Roman calendar which Caesar had ordered to make with the advice of the sages from Egypt and the Christian time reckoning were thus very imprecise.\autocites[We may put aside the general precession of the Earth first found by Hypparchus ca. 127~BCE and now measured to be at about 13 degrees in 1000 years, since this is unrelated to the discussion.][664]{Simon:1994} If the discrepancy was not counted in, the solar calendar lost one day over the course of about 288 years, as it was noticed in the 13th century by John of Sacro Bosco. He remarked that in the 13th century the Julian calendar was actually 10 days ahead of the astronomical calendar. This number can be explained by the difference between 365.25 days of the Julian year and the tropical year of 365.24219 days. This seemed to be the problem of the Egyptian and of the Julian calendars, which held the length of the year to be exactly 365¼ days. The same happened to the calendar of the Moon phases and the projected calculations must have been routinely off the actual phase of the Moon at the expected time on the given day of the solar calendar.\autocites[Compare][398. The problem my article addresses is not the same problem of determining the 30th day of the lunar cycle as outlined in the above-mentioned article, but it is related because the small discrepancy of minutes and seconds could add up to a day over hundreds of years.]{Locher:2006} The case in point here are the calculations that could have been made from the year 1~BCE as the measure of the Moon’s phases across the Julian calendar’s year roll. If we take the actual, astronomical solar year and the lunar month, then, without the imprecision that the human-made calendar introduced, there were several dates when the solar and the lunar rotations coincided exactly. The mathematical formula for that is (‘i’ is the integer number of years):

\[((i\times 365.24219)\mod 29.53) \le 0.25 \notag\]

or

\[(( i\times 365.24219\mod 29.53 \ge 29.27\]

It means a conjunction of the synodic position of the Moon with the position it had in the beginning of the count on the same day of the solar year within ¼ of a day, that is, within 6 hours. During these years the Moon reached it phases (say, a Full Moon) at exactly the same date of the astronomical (but not Julian) solar calendar. These were the years 19 (the foundation of the Easter calendar), 38 (two 19-year cycles), 57, 76, 483, 502, 521, 540, 559, 578 and 598. Let us note that of these dates, 37/38 years and 57 years have been mentioned by Tacitus.

One passage from Tacitus' Annals suggests that in the time of Augustus Romans, probably with the help of the scholars from Alexandria who had earlier helped Caesar to construct a new calendar, became able to sense the cycles of time of the Earth-Moon system that were openly exhibited to the observers in the correlation between the solar and the lunar calendars. Two periods of interest are mentioned in book I of Tacitus' Annals, that of 21 years and of 37 years (Tacitus, Annals, I, 9). These numbers make those who are aware of the astronomical works of Leonard Euler notice that the Romans must have made sense of the shift in the Moon's synodic position. In the years 22 and in 38 from the beginning of any count of time (thus in 21 and 37 full years) the difference between the synodic position of the Moon and both the tropical and sidereal months of the lunar calendar was within one day's precision. In other words, it meant that in 21 and in 37 years from the start of the count there began a year at the end of which the lunar calendar would come almost in sync with itself. In practical terms it meant that the Full Moon on the year 22 CE, coming at March 7th at 1:42, was early by almost two days from the Full Moon on March 8th of the astronomical year “zero” (which has no analogy in the Julian or the proleptic Julian calendar), coming at about 23:17 hours. At the same time, it was late by one day in 38, when it reached the full phase on March 9th at 19:10 PM. If counted from the astronomical year zero, the years 22 and 38 (21 and 37 full years) were those when there was a near conjunction between the solar and the lunar calendar within two days' length. In other words, first the Moon came early by one day of the Julian calendar in 21 full years and then it lost one day against the start of the count.

At the same time, having ruled 57 years, Augustus filled with his life in power a whole cycle of time at the end of which the position of the Moon (the synodic calendar) coincided within 3.5 hours' precision with the solar calendar. In other words, in 21 and 37 years the synodic calendar of the Moon coincided with the time of the Moon's rise and with its taking the position among the stars, while in 57 years the Moon’s phase and the Sun appeared to be almost in a complete sync. It was therefore symbolic that Augustus both acceded to power and died on the same day of August 19th in 57 years from the start of his rulership: this symbolized that the Moon was again reaching its phase in sync with the solar calendar. In other words, this may be considered as the beginning of the Roman eschatological thinking: a day in the solar calendar was first lost, then gained, than the Moon came up even with the Sun, as a human reached the end phase of his life. In other words, this seemed like a whole cycle of life that coincided with the life expectancy periods provided by modern comparative demography methods.

One may thus be certain that after adopting the new calendar with Caesar, the Roman elites immediately started noticing the cycles that were inherent in it and sought to employ them as a measure of time. It is interesting that several numbers in this list (37 and 57 years) suggest that it was not the 19-year cycle that the educated astronomers of Rome used. It was the time when the phase of the Moon coincided with the count of days according to the actual tropical year, and not the Julian calendar.

Let us also notice that there was a long hiatus between the 1st and the 5th century when the Moon and the solar calendar never converged exactly, but in the 5th and 6th century there came critical years when the Sun and the Moon came to move in sync again and made it clearly visible to astronomers and historians.

Let us notice that after that long hiatus the first case of an astronomical conjunction took place in 483 CE. It was also a period in the Easter reckoning after which the lunar calendar needed to be taken back one day. There was only one date in the period between year 1 CE and year 1000 CE that probably allowed the best adjustment in light of the many factors that governed the setting of the date of Easter. The Book of Daniel proposed a period of 483 years as the time to measure the beginning and the end of the repentance period and the time after which a “second coming” of the good times might occur. Let us notice that 483 CE was the only year within 1000-year period when the astronomical cycles of the Sun and the Moon coincided with their original conjunction in the year 1~BCE with one special condition. The lunar calendar added nearly one month by the year 483, but it came short of reaching the same Moon phase as in the year 1~BCE by about 6 hours. As I have shown earlier, in 67 years those keeping track of the solar calendar had to take into account a 6-hour discrepancy: the Julian projection of the calendar dates ran ahead of the astronomical calendar. In other words, if they would consider that the date and time in the Julian calendar had to be pushed back 6 hours in 67 years, or rather, one whole day in 275 years, then they could find a better correlation between the solar date and the lunar epact. Generally, as the Easter tables show, in 483 years the epact for March~21st decreased by one which meant that on the same date of the solar calendar (March 21st, for that matter) the Moon appeared with smaller phase in the days preceding the Full Moon. At the same time it could be imagined that since the epact was less by one day and the Full Moon came later by one day, a new period may have been believed to have started.

One significant mathematical result of using 483 CE as a reference point is paying attention that the difference in days of the Julian lunar calendar (which has 29.5 days for a month) and in weekdays between the year 1 BC or the year preceding the 1st year of the cycle is 5.75 and 1.75, respectively. The mathematical problem was in the fact that the 95-year cycle was not a cycle in the true sense of the term because a computist had to take one day off the lunar month for the cycle to produce the right results. Thus 5 days were taken off the lunar calendar over five 95-year cycles. But what one was to do with other bits of time that remained, that 0.75 of a day? Was it enough of a difference to subtract one epact and make the age of the Moon smaller? The general shift over 483 years was nearly 6 days (5.75). This is where Alcuin’s problem of 798 CE must have originated. If we take these numbers and see which years in the 7000-year cycle correspond to these conditions, we will find that there are 5 dates of these type —-- 483, 2135, 3787, 5439, and 7091. In fact, almost all of them are before the year 6000. Let’s keep in mind that at each of these dates the epact had to be adjusted manually against the normal count of 19-year or 95-year cycles. The difference between the phase of the Moon according to the Julian calendar and its actual showing was 12 hours, meaning that the Moon came earlier. This may explain the problem of early morning Easter celebration: those who were following the Egyptian or Julian calendar had their calculated Moon phase come late, so the Easter would have to be waited for until the morning.

It was this phenomenon that caused the “Easter controversy” of 455~CE. The year 455~CE is known for the disagreement between Pope Leo~I and the Patriarch of Alexandria: the former insisted on an earlier Easter on April 17th and the latter on the Easter on April 24th.\footnote[1376]{ProsperTiro} For the year 455 CE the letter of bishop Proterus of Jerusalem to Pope Leo I mentioned the Easter on XV kalends of May with Moon date 14 on the Easter Sunday.\autocites{Chambert:Protat:2019} Interestingly, Prosper of Aquitaine finished his chronicle in the year 455~CE, which may suggest to us that this year could indeed mean the end of the first cycle from the incarnation of Christ for 5th-century Christians and the beginning of the epoch in which all lunar events would be happening one day later and the Moon would have the lesser phase on the day of Vernal equinox. If we take into account the one-day negative epact shift in 483 years, the 28-year solar calendar’s cycle and the information about Augustus making corrections to the Egyptian calendar between 30~BCE and 23~BCE, the situation becomes clear.\autocites{Skeat:1960}{Skeat:1983}{Skeat:2001} 455 CE was the year 483 from about 28 BCE. Considering modern calculations, we find that the Full Moon on April 17th, 29~BCE, was seconded by the Full Moon on April 18th, 455 CE! Thus the lunar calendar moved back by one day in full 483 years. Let us also remember that Prosper of Aquitaine finished his chronicle in 455~CE, as the the calendar obviously hit a significant snag: if counted from 29~BCE, 455~CE was the 8th year of the 19-year cycle, and so a saltus lunae had to be added: instead, one day had to be taken away from the calendar of Moon phases. It looked as if the calendar lost two days. This is why the letter exchange between Proterus of Alexandria and Pope Leo I was so important: it closed an era stemming from Augustus’ calendar adjustment in Egypt and thus generally brought the timeline of the Roman empire to a new age. Thus in the period of 1000 years the 483rd year was the only one in which the epact and the Easter predictably moved, first by one day back and the second, by 4 days ahead. Thus it was the time of the end and of the new beginning, a possible date for a small-time Apocalypse.

One may also notice that if the length of the tropical year is taken to be equal to 365.24219, 805~CE was the year when the difference between the Julian year and observed Moon’s phases was minimal, just about 30 minutes. Let us keep in mind that the peculiarity of that year was determined by a capitulary that requested scholars to determine the age of the Moon and that led to creation of the so-called “Aachen encyclopedia” of 809~CE. In other words, I propose that the shift of the eschatological end of the 6000 Anno Mundi from ca. 508 CE to 800 CE, made by Gregory of Tours, must have been determined by the proactive calculations that showed a year of total synchronization of the calendar. Let us also remember that 5228 AM was close to a square of 72 years, the period of the precession of the Earth against the fixed stars.

\subsubsection{The Conjunctions of the Astronomical and Civic Calendars}

The abovementioned cycle of about 76 years had a clear explanation in one interesting aspect of the lunisolar calendar. If these discrepancies between the astronomical calendar and the Julian calendar are calculated today with the help of the J. Kepler's and modern data for the parameters of the Earth’s and Moon’ orbits, the actual astronomical conjunction of the Moon’s phase with the solar synodic position came within less than 12 hours as measured against the Julian calendar on approximately the following years: 77, 156, 233, 312, 390, 467, 544, 623, 700, 778. The simplest mathematical formula that is used here is of the following form ($i$ is the integer number of years):

\[\lvert ((i\times 365.24219)\mod 29.53) − ((i\times 365.25)\mod 29.5) \rvert \le 0.5 \notag\]

Its use is justified as the first iteration approach to the problem and is supported by works on astronomy from Leonard Euler on. Again we notice that there are several years, surrounding the “long 5th century”, when the Moon reached its phase at about the same hour as in the year of the birth of Christ (but not necessarily on the same date of the Egypian/Julian calendar). Remarkably, one may find on this list the year 467 CE, when after the blood Moon of 462 Hydatius finished his Chronicle by describing the rise of Euric to power and the failed expedition of emperors Leo and Anthemius, finally disappearing from he scene by 468~CE.\autocites[21]{Wieser:2019} Let us also note that the famous Late Antique epistolographer Sidonius Apollinaris effectively wrapped up his ascending career that had made him the prefect of Rome by 468~CE and was appointed the bishop of Clermont in 469~CE.\autocites[xxviii-xxxi]{SidoniusApollinaris:1915:1} In other words, it seems that astronomers and people in power took notice of the time when the astronomical conjunction between the Moon's phase and the solar calendar was supported by the calculations made on the basis of the Julian calendar and the 29.5 day lunar month. As I will show later, this feeling may have been exacerbated by other astronomical phenomena that made people sense the influence of the cormor on their daily lives.


\subsubsection{The Conjunctions of the Astronomical and Civic Calendars within Hours' Precision}

In some cases the problem was less in the actual count of days since in the Ancient world the astronomers managed to create calendars that helped intercalate any extra days into the year (with an embolistic month and a saltus lunae). The problem lay in the discrepancy in hours between the actual astronomical calendar and the calculated Julian calendar, which was imprecise even in its best representations. Although the difference in days could be significant, the correlation between the actual time of the Moon reaching its phase and the projected value was within one hour for the following dates.

\[\{\lvert ((i\times 365.24219)\mod 29.53) - ((i\times 365.25)\mod 29.5) \rvert\} \le 0.5 \notag\]

We will notice that the years 223 CE (a year that was made significant in the Chronicle of Julius Africanus in the start of the Easter time reckoning in 222 CE, the 13th year in the Easter cycle,\autocites{Africanus:2007} 359 CE (the last good year of Julian the Apostate’s reign when he at last managed to reunite the empire under the power of a single ruler), 392 CE (the year when the peace and unity in the Empire were restored with the accession of Theodosius I), and 417 CE (the year when the Goths were settled in Aquitaine, thus ending their incessant wanderings) were such when the projected cycles of the Moon’s phases coincided with the observed sightings within two hours or less. In 475, however, the Moon was late by almost one day against the calculated values.

\subsubsection{The Draconic and the Anomalistic Month’s Cycles}

Due to the complexities of the lunar motions there were several types of the lunar month —sidereal, anomalistic, and draconic. We can consider when the motion of the Moon according to the regular phase month of 29.53 days coincided with its elevation or the time of rise during the period of up to 600 years. We are seeking the years when the synodic position of the Moon on a given date correlated with its visible size, inclination, or angular position over the horizon. Considering that many astronomers and specialists in chronology in Late Antique world counted time from 8 BCE, we will get several dates of historical importance. Considering as a basis the anomalistic month of 27.554 days (which is directly related to the visible size of the Moon and originates from the Moon orbit’s precession around the Earth), we get some meaningful dates only if we count from 8 BCE. But, since 8 BCE was 746 “ab urbe condita,” it was the 10th showing of the {``}large Moon{''} due to the cycle of the Moon's elliptical orbit that made the Moon appear in exactly the same perigee spot close to the Earth every 74.6 years. Thus we can find among those years when the Moon was in the same spot, of the same phase and size as in the year 1 BCE the years 337 CE (the death of Constantine the Great) and 496 CE (the baptism of Clovis).

The simplest mathematical formula that is used here is of the following form:

\[\lvert ((i\times 365.24219)\mod 29.53) − ((i\times 365.24219)\mod 27.554) \rvert \le 1 \notag\]

In other words, the founding of Rome, the death of the first Christian emperor and the first catholic baptism of a new ruler, a barbarian king, all happened with the Moon’s positions having approximately the same parameters except for the fact that Moon was in its apogee and thus seemed having the smallest possible size. Let us notice upfront that in 455 the Moon had the same largest size as it did at the moment of establishing Rome in 753 BCE.

At the same time, we may consider the draconic month of 27.212 days, which corresponds to the Moon's orbit plane wobbling and the Moon reaching the same visual latitude in the sky (it visually meant the Moon being at exactly the same angle above the horizon).

\[\lvert ((i\times 365.24219)\mod 29.53) − ((i\times 365.24219)\mod 27.212) \rvert \le 1 \notag\]

Considering the fact that the initial position of the Moon was taken note of between 8 BCE and 5 CE, we may also get several interesting historical dates: 285 CE (if the initial latitude of the Moon's position was observed at 4 CE), the creation of Tetrarchy, 417 CE (the settlement of Visigoths in Aquitaine, with the initial observation at 5 CE), and 412 CE (with 1 BCE as the first observation date), the year when the observed phase of the Moon and the one calculated according to the Egyptian calendar coincided. At the same time 412 CE was the year when, after Alaric II had died in Rome, Athaulf married Galla Placidia and left. In other words, one may surmise that between the very end of the 4th century (from about 275 CE) to the beginning of the 6th century (to about 504 CE), the visual aspects of the Moon's position and shape began to repeat those that were visible at around the birth of Christ.

Thus one could obviously see in the motions of the Moon two patterns: one, a significant regularity of about the length of the maximum length of human life that was accompanied by the precession of the Earth against the fixed stars and two, a series of cycles of synchronism between the Sun’s and the Moon’s motions that were hard to calculate at the moment.

Let us summarize the years over the first five centuries of Christian history in a table. The years of the generally known historical significance are highlighted.

\begin{tabular}{| c | c |}

\hline

\thead{Type of Conjunction} & \thead{Years}  \\

\hline

\thead{Astronomical \\ Conjunctions} & 19, \textbf{38}, \textbf{57}, 76, \textbf{483}, 502, 521, 540, 559, 578, 598 \\
\thead{Astronomical \& Civic \\ Calendar Conjunctions} & 77, 156, 233, \textbf{312}, \textbf{390}, \textbf{467}, 544, 623, 700, 778 \\
\thead{Time of Day \\ Conjunctions} & \textbf{223}, 359, \textbf{392}, \textbf{417}, 475 \\
\thead{Draconic Month \\ Conjunctions} &  \textbf{285 CE} (from 4 CE), \textbf{417 CE} (from 5 CE), \textbf{412} \\
\thead{Anomalistic Month} & \textbf{337}, \textbf{496} \\

\hline

\end{tabular}

\subsubsection{Supermoons}

Among the various cycles that could be heuristically constructed by Late Antique astronomers one needs to consider that of the Moon's perigee and apogee, which was exhibited in the Moon's showing larger or smaller. Modern data that was acquired with the help of laser measurement for the years 2001-2022 suggests that the Moon reaches its perigee about every 398 days, although this figure is more likely to be within the range of 386 and 414. Let us approach the problem of how Sepermoons or Micromoons might have showed up in historical sources from 2 different perspectives. As modern measuruments suggest, the cycle of the perigees and apogees do not fully coincide with the synodic lunar month or draconic month. The periods of supermoons come in groups of 3 to 4 months, followed by the periods of micromoons. The period between the start of one period of supermoons and the start of the next one is highly variable and can take the length of 386 to 414 days. It happens because of the precession of the lunar orbit and of the variation of the 2 main axes of its ellipse. Since the length of the anomalistic month is 27.554 days, the period of the perigees seems to lie between 14 and 15 of these months, i. e., between 385.76 and 413.31 days. The anomalistic month as the time between the Moon appearing at the same visible diameter is effectively the other way of communicating the visible size of the Moon.

Thus it is viable to test 2 approaches of measuring the correlation between the cycles of supermoons and the more traditional ways of the synodic lunar calendar of phases. The supermoons can be measured in their relationship to the solar calendar and in relationship to the Moon's synodic month of 29.53 days. On the one hand, we will take various values of the length of time between the cycles of perigees, and will try to calculate on what days of the solar year they fall. We will limit the results to the cases when the Supermoons or Micromoons fall on nearly exactly the same date of the solar year as it happened at the beginning of the count. For the purposes of finding a correlation between this phenomenon and events in Late Antique and medieval history let us calculate the years in which the Moon would appear the largest, but let us at first limit these calculations to the count of days of the solar calendar. In other words, let us first find out the years when the Moon would be in the perigee on the same day of the solar calendar. At this moment we will not consider the phase of the Moon since it makes the calculations more complex and the results much less interesting from the point of view of a historian.

For this purpose let us construct a formula for the values of the number of years ($years$) and the number of days in the perigee cycle ($PerigeeCycleLength$), of which the later will be considered as a range. Since the observation of the larger Moon image may vary by several days, let us consider the limits on the values in the following manner:

\small

\[((365.24219 \times years) \mod PerigeeCycleLengh) \le 2 \notag\]

or

\[\lvert PerigeeCycleLength - ((365.24219 \times years) \mod\ PerigeeCycleLengh) \rvert \le 2 \notag\]

\normalsize

The second condition is needed to take into account those cases in which the remainder of the days is very close to the perigee cycle length, which means that the end of a cycle would have been reached in 2 days.

In other words, we will be looking at the years when the Moon would appear on the same Julian calendar date plus or minus two days. For the perigee period between 396 and 398 days (which one can take as an approximation from the modern data tables), there are several historically important dates: for 397 days the Moon seems to be of the same size as in the year 1 BCE, the years of Jesus' birth (of currently unknown phase), in the year 800 CE, the year of the coronation of Charlemagne. For the period of 398 days we get the synchronization of the Moon's size in 814 CE, the year when Charlemagne passed away, and 1215 CE, the year when the Magna Carta was signed by king John of England and the rebellious barons. For the period of 396 days the year 1030 CE comes up, the year when Ademar of Chabannes wrote his Chronicle in the wake of the expecations of the world coming to an end. This period also gives the year 401 CE. If the period is 396 days, one gets 412 CE as the year of the conjunction. In other words, even without knowing the Moon's phase at the perigee and the apogee we come across 4 critically important dates in history. This suggests that the Moon's position in the same point of its orbit, perigee or apogee, might have indeed influenced rulers' and historians' perception of the current events and possibly made them mark that year as significant.

On the other hand, we will construct a correlation between the Supermoons (with its close relationship with the anomalistic month) and the synodic month, in this case trying to find when the Supermoons indeed fell on the time of the Full Moon within a reasonably small margin (a day or less). We will the correlate these results with the particular years, and limit the results to only those years when the Moon happened to be in perigee during its synodic Full phase it the time that was considered that of religious festivals at the particular time of the year. In other words, we will be interested in Supermoons that fall on 1. the days of the birth of Christ, Christmas 2. Easter 3. the beginning of the Constantine's civil year at September 1st.

If we would look for the Full Supermoon falling within 2 days of the original one, we get a cycle of 8 years. At the same time, if we further limit ourselves to determining when the Supermoon really happened during the same phase, we get the following, much more restrictive set of requirements. The first is

\[PerigeeCycleLength \times PerigeeCycles \mod 29.53 \le 1 \notag\]

It will give us the number of perigee cycles, not Julian years. To get the number of Julian years we will need to mutiply that number of the perigee cycles by the factor of the length of the solar year (365.24219) divided by the length of the perigee cycle. Moreover, since the spread of the Full Supermoons will fall on practically every month of the year, which hardly interested those who were keeping the calendar, we need to limit the results to the years when the Full Supermoon (or whatever it was in the year 1 BCE) fell within the same month, as it did in the original conditions. Thus we must first calculate the number of Julian years from the given number of perigee cycles, and then limit ourselves to the years when the Full Supermoon (or other phase) fell within one month's duration to the original date of the first sighting, that is, 1 BCE.

Let us remember, though, that we have already put a limit on the perigee date so that it would happen within one day of the Lunar phase. In other words, we look for years when the Full Supermoon (or whatever other phase there was in 1 BCE) fell close enough (within 1 day) to the orginal date of the first of three showings of the Full Supermoon.

This way we get a number of interesting results. The synchronization of that kind is very rare and in the given conditions (with the perigee cycle counted as being of the length of between 396 and 398 days) it provides few dates. This happens because of the very stringent conditions we have placed on the synchronicity of the Full Supermoon. But these dates also have some relationship to the chronological schemes of the past: for example, if the perigee cycle is taken to be 397 days, the Full Supermoon will happen in 1000 years! This is where the millenarian considerations may have come from. In the case of the cycle of 396 days the synchronizations do not extend beyond 712 CE. And in the case of 398 days there are Full Supermoons in 702 CE, 847 CE and 1137 CE: the first being the date when computists began considering putting together a new Easter cycle table, and the last being the date of the nuptial union between the dynasty of Capetians (Louis VII) and that of the Dukes of Aquitaine (Eleanor). This nuptial union exemplified the rare juncture of the solar and the lunar calendars. There are also two dates related to Late Antiquity: 402 CE in the case of 397-day cycle and 412 CE in the case of the 398-day cycle.

In other words, we will be looking at the years when the Moon would appear on the same Julian calendar date plus or minus two days. For the perigee period between 396 and 398 days (which one can take as an approximation from the modern data tables), there are several historically important dates: for 397 days the Moon seems to be of the same size as in the year 1 BCE, the years of Jesus' birth (of currently unknown phase), in the year 800 CE, the year of the coronation of Charlemagne. For the period of 398 days we get the synchronization of the Moon's size in 814 CE, the year when Charlemagne passed away, and 1215 CE, the year when the Magna Carta was signed by king John of England and the rebellious barons. For the period of 396 days the year 1030 CE comes up, the year when Ademar of Chabannes wrote his Chronicle in the wake of the expecations of the world coming to an end. This period also gives the year 401 CE. If the period is 396 days, one gets 412 CE as the year of the conjunction. In other words, even without knowing the Moon's phase at the perigee and the apogee we come across 4 critically important dates in history. This suggests that the Moon's position in the same point of its orbit, perigee or apogee, might have indeed influenced rulers' and historians' perception of the current events and possibly made them mark that year as significant.

\begin{tabular}{| c | c | c | }

\hline

\thead{cycle length, days} & \thead{Moon perigee cycles' \\ conjunctions with solar calendar} & \thead{Full Supermoons \\ phase conjunctions \\ with the solar calendar}\\

\hline

386 & & \textbf{79}, \textbf{410} \\

389 & & \textbf{417}, 445, 466 \\

392 & & 407, 444, 454 \\

395 & & 754, 776, 855, 870 \\

396 & 401, 412, \textbf{1030} & \\

397 & \textbf{800} & \textbf{410}, \textbf{1000} \\

398 & \textbf{486}, \textbf{814}, \textbf{1215} & 402, 412, 1137 \\

405 & & 455 \\

406 & & \textbf{800} \\

407 & & \textbf{813}, \textbf{833} \\

408 & & \textbf{751} \\

411 & & \textbf{813} \\

412 & & 772, \textbf{811} \\

413 & \textbf{467} (Hydatius, Sidonius) & \\

416 & 764 & \\

\hline

\end{tabular}

\section{Conclusions}

The lunisolar calendar that Christians used for the calculation of the main holiday of their religion thus required a lot of skills which the specialists in calendar were gaining through all Late Antique history and up to the Carolingian age. The first episode of facing off with the irregularities of the lunar calendar which eventually stemmed from the harmonics of the Moon’s precession on its orbit around the Earth came at about 410~CE when the problem did make historians wonder whether the Heavens all went against the normal course of time. But since the matter was in that one day which had to be subtracted, historians and specialists in the calendar had to wait until the year 483 when some operation with the adjustment of the count may have been expected. Once the necessary manipulations were made or confirmed ca. 455~CE (and it may have been the first adjustments of the calendar since 222~CE), the count of the lunar month’s days was restored to its normal path and chronographers could rely on Easter tables as the source of correct lunar sightings.

Let us notice that scholars who were actively involved in the construction of the Christian history's chronology were limited in their choices by the astronomical peculiarities of the Earth-Moon system. The total conjunctions of the astronomical Solar and Lunar calendars took place, some within the 1st century CE, and the next one, in 483 CE. This was also a special year because the lunar calendar lost one day. Thus the 5th century was the time of heightned expectations of whether the calendar and the Moon's showings will repeat those that accompanied the birth of Jesus. The Full Supermoon (or whatever phase it was on December 25th, 1 BCE) may have repeated in 410 CE (the entry of Goths into Rome), in 467 CE and in 476 CE (the Fall of the Roman Empire), marking the coming of the time very similar to Jesus' birth. The Full Moon was supposed to repeat December 25th, 800 CE and in the year 1000 CE. This may have determined the setting of the biblical calendar in a way that put the birth of Christ on 5199 CE (making the year 800 CE, the year of the Full Supermoon or of its phase on December 25th, 1 BCE) a critical turning point.

This suggests an answer to the question emerging from the historiography: was the Christian lunisolar calendar important to more than select few? Studies of P.~Brownw L.~Pietri, and R.~Van~Dam have shown the importance of the holy men, of the ascetics and bishops, in making an example for their communities of living according to the principles of the Christian lunisolar calendar. Their attention to conforming with the lunisolar cycle in their own actions and in their emphasis on the correlations between this calendar and the events in history suggested that astronomy did influence their and the rulers' behavior as models for the masses.

\printbibliography

\end{document}